\documentstyle[12pt]{article}
\newcommand{\la}{\langle}
\newcommand{\ra}{\rangle}
\begin{document}
%
\title{
Analyses of multiplicity distributions of $e^+e^-$ and
$e$-$p$ collisions by means of modified negative
binomial distribution and Laguerre-type distribution:
Interrelation of solutions in stochastic processes}
\author{ Minoru Biyajima$^{1}$, Takeshi Osada$^{2}$
and Kenji Takei$^{1}$ \\
$^{1}$Department of Physics, Shinshu University, 
Matsumoto 390, Japan \\
${}^{2}$Department of Physics, Tohoku University, 
Sendai 980-77, Japan }
\date{\today}
\maketitle
 
\begin{abstract}
A pure birth stochastic process with several 
initial conditions is considered.  
We analyze multiplicity distributions of $e^+e^-$
collisions and $e$-$p$ collisions, usig
the  Modified Negative Binomial Distribution
(MNBD) and the Laguerre-type distribution. 
Several multiplicity distributions show the same 
minimum $\chi^2$'s values in analyses by means of two 
formulas: In these cases,  we find that a parameter $N$ 
contained in the MNBD
becomes to be large. Taking large $N$ limit in the MNBD, 
we find that the Laguerre-type distribution can be derived from it. 
Moreover,  from the generalized MNBD we can 
also derive the generalized Glauber-Lachs formula.
 
Finally stochastic properties of QCD and multiparticle dynamics are
discussed.
\vspace*{0.5cm}
\noindent
 
PACS  number(s): 13.65.+1, 34.10.+x \\
\end{abstract}

\noindent
{\bf 1. Introduction}~~\\
The Negative Binomial Distribution (NBD)
played the important role in analyses of multiplicity
distributions in high energy collisions\cite{biya83,BIYA-SUZU}.
 However, to resolve the problem concerning 
a parameter $k$ of the NBD the following
scheme adopted from QCD was proposed in Refs. 
\cite{durand86,hwa86}. 
In QCD we can consider a pure gluon
bremsstrahlung process, $ g \to g + g$. The evolution equation is
\begin{equation}
   \frac{\partial P(n, t)}{\partial t} = -\lambda n P(n, t) +
\lambda(n-1)P(n-1, t),\label{eq:1}
\end{equation}
where $P(n, t)$, $ t \propto ln[ln(Q^2/\mu^2)/ln(Q_0^2/\mu^2)]$,
and  $\lambda = N_c$(color factor)$/\varepsilon$(cutoff)
are the gluon distribution,
a proper variable in QCD, and the production
rate of gluons, respectively.
Assuming $n_0$ gluons at $t = 0$, i.e., $\delta_{n,n_{0}}$, they
obtained the following distribution with $ \bar{n}(t) =
n_{0} exp(\lambda t) $,
\begin{equation}
   P_{n_0}(n, t) =\frac{(n-1)!}{(n-n_{0})!(n_{0}-1)!}
(\frac{n_{0}}{\bar{n}})^{n_{0}}(1-\frac{n_{0}}{\bar{n}})^{n-n_{0}}.
\end{equation}
Using the local parton duality between QCD and multiparticle 
dynamics Eq.(2) was applied to the data of high energy 
collisions in Refs. \cite{durand86,hwa86}.
However, Eq.(2) does not show satisfactory results, because of the 
fixed multiplicity at $t = 0$.  This suggests that the fluctuation 
at $ t = 0$ is important.
 
Later to improve the defects of Eq. (2),
 two initial conditions, the Poissonian distribution and 
the binomial 
distribution have been considered: Using Eq. (1) with them, 
the Lagueree-type distribution in Ref. \cite{bks87} 
and the Modified Nagative Binomial Distribution (MNBD)
in Refs. \cite{chi90}-\cite{ONBS97} have been derived.
See also Refs.\cite{fink88}-\cite{chau92}.\\
 
In the present paper we analyze the data
of  $e^+e^-$ collisions including the data of $e$-$p$ collisions
in Refs.\cite{HRS}-\cite{h1} by means of three formulas.
Several results show that the
minimum $\chi^2$ values by the MNBD with large $N$ contained in
the formula are  coincided with those by the Laguerre-type
distribution.  See Table I.
Thus we investigate whether or not there is any
interrelation between the MNBD and the Laguerre-type 
distribution.
 
Contents of this paper are as follows:  In the next
paragraph and the third one,  we introduce the MNBD and the
Laguerre-type distribution, respectively.  In the fourth one,
we  consider the large $N$ limit of the MNBD. In the fifth one,
we consider the generalized MNBD.  Finally our concluding
remarks are presented.\\
 
\vspace*{0.3cm}
 
\noindent
{\bf 2.  MNBD and its application to data of $e^+e^-$ and
$e$-$p$ collisions}~~\\
 
We assume that the solution of Eq.(1), i.e., $ P(n,t)$,  is 
useful for description of multiplicity, i.e.,
in multiparticle dynamics, using the local parton duality.
The MNBD is obtained from Eq.(1), using the following 
initial condition,
 
\begin{eqnarray}
P(n,t=0)
=  {}_{N}C_n \alpha^n(1-\alpha)^{N-n},
\end{eqnarray}
where $N$  is an integer and $N\alpha$ is the average
number of clusters. The generating function of the MNBD is
\begin{equation}
	\Pi(u) = \sum_{n=0}^{}P(n)u^{n}
	       = [1-r_1(u-1)]^{N}[1-r_2(u-1)]^{-N},
	\label{eq:0}
\end{equation}
where $ r_{1} = p - \alpha (p + 1) $, and $ r_{2} = p $ with
$ p = \exp( \lambda t) - 1. $  In this case $\lambda$ is the
production rate of hadrons. 
%
 
The MNBD is given by
the following function \cite{chi90,NBS}, neglecting
the second variable,
\begin{eqnarray}
  P(0) &=& \left[ \frac{1+r_{1}}{1+r_{2}} \right]^N, \nonumber \\
  P(n)&=&\frac{1}{n!} \Big[ \frac{1+r_1}{1+r_2} \Big]^N
         \sum_{s=0}^{n}{}_{n}C_{s}~
         \Big[ (-A)^s \frac{N!}{(N-s)!}   \Big] \nonumber \\
       && \times \Big[~B^{n-s}~\frac{(N+n-s-1)!}{(N-1)!}~\Big],
      \label{eq:05}
\end{eqnarray}
where $A=1+r_1/(1+r_1)$ and $B=r_2/(1+r_2)$.
  The $r_{1}$
and $r_{2}$ are expressed by the average multiplicity $\langle 
n \rangle$ 
and the second moment $C_{2} =
\langle n ^2 \rangle / \langle n \rangle ^2.$  In this case
$r_{1}$ is real and $r_{2}>0 $
\begin{eqnarray}
r_{1} =\frac{1}{2} \left( C_{2}-1-\frac{1}{N} \right)
\langle n \rangle -\frac{1}{2}, \quad
r_{2}
= r_{1} + \frac{\langle n \rangle}{N}. \label{eq:3}
\end{eqnarray}
%
Our analyses of the charged multiplicity
distributions in $e^+e^-$ and  $e$-$p$ collisions are shown
in Table I. 
 
The KNO scaling function of the MNBD has been calculated
by the usual procedure, $\displaystyle{ z =\!\!\lim_{n,\langle n
   \rangle\rightarrow \infty}}\!\! n/\langle n \rangle$ fixed,
in a recent paper \cite{ONBS97}:
\begin{eqnarray}
     \Psi(z)\equiv \lim_{n,\langle n \rangle\rightarrow
     \infty}\langle n \rangle P(n)
                =\left(\frac{r'_1}{r'_2}\right)^{N}
             e^{-\frac{\langle n \rangle}{r'_2}z}
             \sum_{j=1}^{N}{}_NC_{j} \frac{1}{\Gamma(j)}
             \left(\frac{r'_2-r'_1}{r'_1}\right)^{j}
   \left(\frac{\langle n \rangle}{r'_2} \right)^{j}z^{j-1}.
        \label{eq:5}
\end{eqnarray}
The parameters $r'_1$ and $r'_2$ in Eq. (\ref{eq:5}) are
given by
\begin{eqnarray}
  r'_{1} &=&
 \frac{1}{2} \left(C_{2}-1-\frac{1}{N} \right)\langle n 
  \rangle,\quad
  r'_2
         = r'_1 + \frac{\langle n \rangle}{N},
 \label{eq:06}
\end{eqnarray}
because $\la n \ra \gg 1$.
Our results by the KNO scaling functions are given in 
Table II.\\
 
\noindent
{\bf 3. Laguerre-type distribution and its application
to data}\\

Using Eq. (1) with the
Poissonian distribution as the initial condition,
\begin{equation}
P(n,t=0)
=  \delta_{n,k} \frac{\langle m \rangle^k}{k!}
\exp(-\langle m \rangle),\\
%
%
\end{equation}
we obtain the Laguerre-type distribution.
(Notice that $\langle m \rangle = 2/( C_2 -1 +1/{\langle
n \rangle})$ has been interpreted as the
number of clusters at $t=0$.)
\begin{eqnarray}
&&P(0) = \exp(- \langle m \rangle ),\nonumber\\
&& P(n) =\langle m \rangle ^2 \exp(-{\langle m \rangle} )
 \frac{ ( \langle n \rangle - \langle m \rangle) ^{n-1}}
{ \langle n \rangle^n }~\frac{1}{n}~L^{(1)}_{(n-1)}
(- \frac{ \langle m \rangle^2}{ \langle n \rangle
- \langle m \rangle }), \quad n = 1,2,... \quad \label{eq:62}
\end{eqnarray}
It should be stressed that Eq.(10) was derived in various
approaches and applied to analyses of data in high energies
in Refs. \cite{fink88,gupta92,chau92}.
 
The KNO scaling function of Eq.(\ref{eq:62})
is  given as 
\begin{eqnarray}
\Psi(z)
     &=& \frac{ {\langle m \rangle}_k }{\sqrt{z}} \exp[
     -{\langle m \rangle}_k(z + 1)]
 \times I_{1}( 2{\langle m \rangle}_k \sqrt{z}) \label{eq:11},
\end{eqnarray}
where $\langle m \rangle_k=2/(C_2-1)$.
Our results by means
Eqs. (10) and (11) are also given in Table I and II.
Therein we can see that several minimum $\chi^2$'s values
by Eq.(5) with large $N$
are coincided with those by Eq.(10).
This fact suggests us that the MNBD in large $N$
limit probably contains the Laguerre-type distribution.
Thus in the next paragraph we investigate this 
 possibility.\\
 
\noindent
{\bf 4. Large $N$ limit of the MNBD}~~\\
 
We take the large $N$ limit in Eq. (\ref{eq:05}):
\begin{eqnarray}
%
(\frac{1+r_1}{1+r_2})^N
\stackrel{\mbox{\scriptsize{\it large N}}}{\longrightarrow}
\exp( - \frac{2}{C_2 - 1+1/\langle n \rangle})
 = \exp (- {\langle m \rangle})
%
%
%
\end{eqnarray}
%
Eq. (12) is interpreted as the distribution of
clusters at the initial time $t = 0 $ as
Ref. \cite{bks87}.
 The origin is attributed to the Poissonian distribution at
 $t = 0$, which can be reduced from the
binomial distribution as
\begin{equation}
 _{N}C_n \alpha^n (1 - \alpha)^{N - n}~~
\stackrel{\mbox{\scriptsize{\it large N}}}{\longrightarrow}~~
 \frac{(N\alpha)^n}{n!} \exp(- N\alpha).
\end{equation}
Using $N\alpha = \langle n \rangle /(1 + r_2) = \langle
m \rangle $,  we obtain Eq.(10) as Ref. \cite{bks87}.
%
Using an approximation ${}_{N}C_j \approx N^j/j!$,
%
%
%
we obtain Eq.(10) neglecting the contribution with 
${\cal O}(1/N)$. Thus it
%
%
%
it can be said that Eq.(5) contains Eq.(10).  
For the KNO scaling functions, we have Eq.(\ref{eq:11})
taking the large $ N$ limit in Eq.(17),
and using the following approximation
\begin{eqnarray}
\lim_{N\to \infty}
\frac{1}{N}L_{N-1}^{(1)}\Big(
\frac{-1}{N}\langle m\rangle^2_k z \Big)\!\!&\approx&\!\!
\frac{1}{\langle m \rangle_k\sqrt{z}}~I_1(2\langle m 
\rangle_k\surd{z}).
\end{eqnarray}
The present calculation shows that there is an interesting
interrelation between solutions of Eq (1).
Here it is worthwhile to notice that a similar interrelation 
between the Ising model with the free lattice and Eq.(10) was 
used in Ref.[12]. \\
 
\noindent
{\bf 5. Generalized MNBD}~~\\
According to calculations of Refs. [2,3], we can consider
Eq.(1) with the immigration as follows
\begin{equation}
\partial P(n,t)/\partial t = -\lambda n P(n,t) +
\lambda(n-1)P(n-1,t)
             -\lambda_0 P(n,t)+\lambda_0 P(n-1,t),
\end{equation}
where $\lambda_0$ is an immigration rate 
corresponding to $q\to q +g$ process in QCD[2-3]. 
(On the other hand, in multiparticle dynamics the physical 
meaning of $\lambda_0$ may be interpreted as 
possible contributions from neutral excited hadrons.)

By making use the
same initial condition, Eq. (3), we obtain the following
generating function with $k=\lambda_0/\lambda$ as
\begin{equation}
	\Pi(u) = \sum_{n=0}^{}P(n)u^{n}
    = [1-\tilde{r}_1(u-1)]^{N}[1-\tilde{r}_2(u-1)]^{-k-N},
\end{equation}
where
\begin{eqnarray}
  \tilde{r}_{1}=\frac{\langle n \rangle}{k}\left \{ 1 -
 \sqrt{\frac{k+N}{N}
   \left[ 1 - k \left( C_2 -1-\frac{1}{\langle n \rangle}
\right)
         \right] }~~\right \}, \quad
   \tilde{r}_{2}= \frac{N \tilde{r}_1 + \langle n
\rangle}{k+N}. \nonumber
\end{eqnarray}
Notice that Eq. (16) contains the NBD($N=0$) and
the MNBD($k=0$). 
The probability distribution is obtained from $\Pi(u)$ as
 \begin{eqnarray}
  P(0) &=& \Pi (0) = \frac{\left(1+\tilde{r}_{1} \right)^N}
{\left
(1+\tilde{r}_{2} \right)^{N+k}},
                        \nonumber \\
  P(n)&=& \frac{1}{n!}\frac{(1+\tilde{r}_1)^N}
{(1+\tilde{r}_2)^{N+k}}
          \sum_{s=0}^{n}~{}_{n}C_{s}
    \Big[~(-\tilde{A})^s \frac{N!}{(N-s)!}~\big]  \nonumber \\
        && \times \Big[~\tilde{B}^{n-s} \frac{(N+k+n-s-1)!}
{(N+k-1)!}~\Big]
\end{eqnarray}
where $\tilde{A}=\tilde{r}_1/(1+\tilde{r}_1)$ and
      $\tilde{B}=\tilde{r}_2/(1+\tilde{r}_2)$.
 
We obtain the following generalized Glauber-Lachs formula
in Ref.[1] taking large $N$ limit in Eq.(17), 
\begin{eqnarray}
\mbox{ Eq. (17)}
\stackrel{\mbox{\scriptsize{\it large N}}}{\longrightarrow}
P(n)  &=&  \frac{(p \langle n \rangle/k)^n}{(1+ p \langle n
           \rangle/k)^{n+k}}
      \exp[ -\frac{ \langle n \rangle}{(1 + p \langle n
           \rangle/k)}] \nonumber \\
      &&\hspace*{0cm}  \times
      L_n^{(k-1)}(-\frac{k/p}{(1 + p \langle n \rangle/k)}),\\
&& \hspace*{-7.0cm}\mbox{where} \qquad \qquad \hat{r}_{1}
\hspace{0cm} \stackrel{\mbox{\scriptsize{\it large N}}}
{\longrightarrow}
\frac{\langle n \rangle}{k}\left \{ 1 -
\sqrt{
       1 - k \left( C_2 -1 -1/ \langle n \rangle \right)
  }~~\right \} \equiv \frac{\langle n \rangle}{k}p,\nonumber
 \mbox{and}  \hat{r}_{2} - \hat{r}_{1} = 
\frac{\langle n \rangle}{N}.\nonumber
\end{eqnarray}
\\ 
 
\noindent
{\bf 6. Concluding remarks}~~\\

Equations (5) and (6) and the NBD are used in analyses of the data. 
We have found that several minimum $\chi^2$ values are coincided 
with each other. 
Referring this fact, we have shown that Eq.(10) is derived from 
Eq.(5) as $N$ is large. Moreover the generalized Glauber-Luchs 
formula is also derived from Eq.(17) as $N$ is large.
 
We consider physical correspondences between QCD and multiparticle
dynamics in Eq.(1). Using Eq.(4), $\exp(\lambda t)=1+r_2$,
$\lambda$ and $t$ can be estimated by the data. They are shown in
Fig. 2 and Table III. 
We obtain an interesting expressions for $\lambda$ and $t$ in
multiparticle dynamics.  The different interval concerning $t$ is 
probably  attributed to the hadronization. More data at 
$\sqrt{s}$=135 GeV are necessary to confirm them.
 Finally it should be noticed that there is a different approach to 
stochastic equation in QCD in Ref. [22].

\vspace*{0.35cm}
 
\noindent
{\bf Acknowledgments:}~~
One of the authors (M. B.) is partially supported by the
Grant-in Aid for Scientific Research from the Ministry 
of Education, Science, Sport and
Culture (No. 06640383) and  (No. 09440103).

\clearpage
\noindent
{\large Table captions}\\
 
\noindent
{\bf Table I:}\\
The minimum values of $\chi^2$'s for fitting of discrete distribution
to the experimental data using the MNBD the NBD and the
Laguerre-type distribution, Eq.(10),
respectively. The data of $\langle n \rangle$
and $C_2$ are used. `$\ast$' denotes value of the minimum $\chi^2$~'s.
The NBD is given by $P(n)=\Gamma(n+k)/\Gamma(n+1)~\Gamma(k)
~(k/\langle n \rangle)^k~(1+k/\langle n \rangle)^{-(n+k)}$.\\
 
\noindent
{\bf Table II:}\\
The same as Table I but for fitting of the KNO scaling function
obtained from the MNBD, Eq. (7), the gamma distribution
and the Laguerre-type distribution, Eq.(11),
respectively. The gamma distribution is given by
$\Psi(z)=k^k/\Gamma(k)~e^{-kz}~z^{k-1}$.\\
 
\noindent 
{\bf Table III:}\\
Physical correspondences between $\lambda$ and $t$ in Eq.(1) in
QCD and mutiparticle dynamics.\\
 
\noindent
{\large Figure Captions}\\
 
\noindent
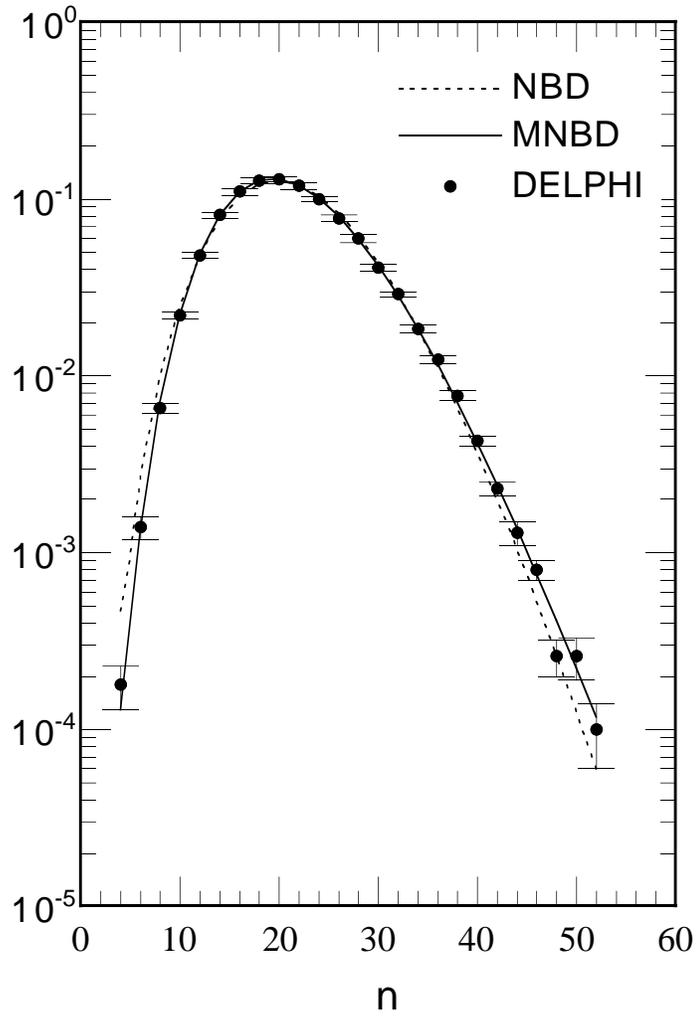
{\bf Fig.1:} \\
~The result of fitting by NBD and MNBD to the data observed in
$e^+e^-$ collisions by DELPHI Collab.[14]. \\

\noindent
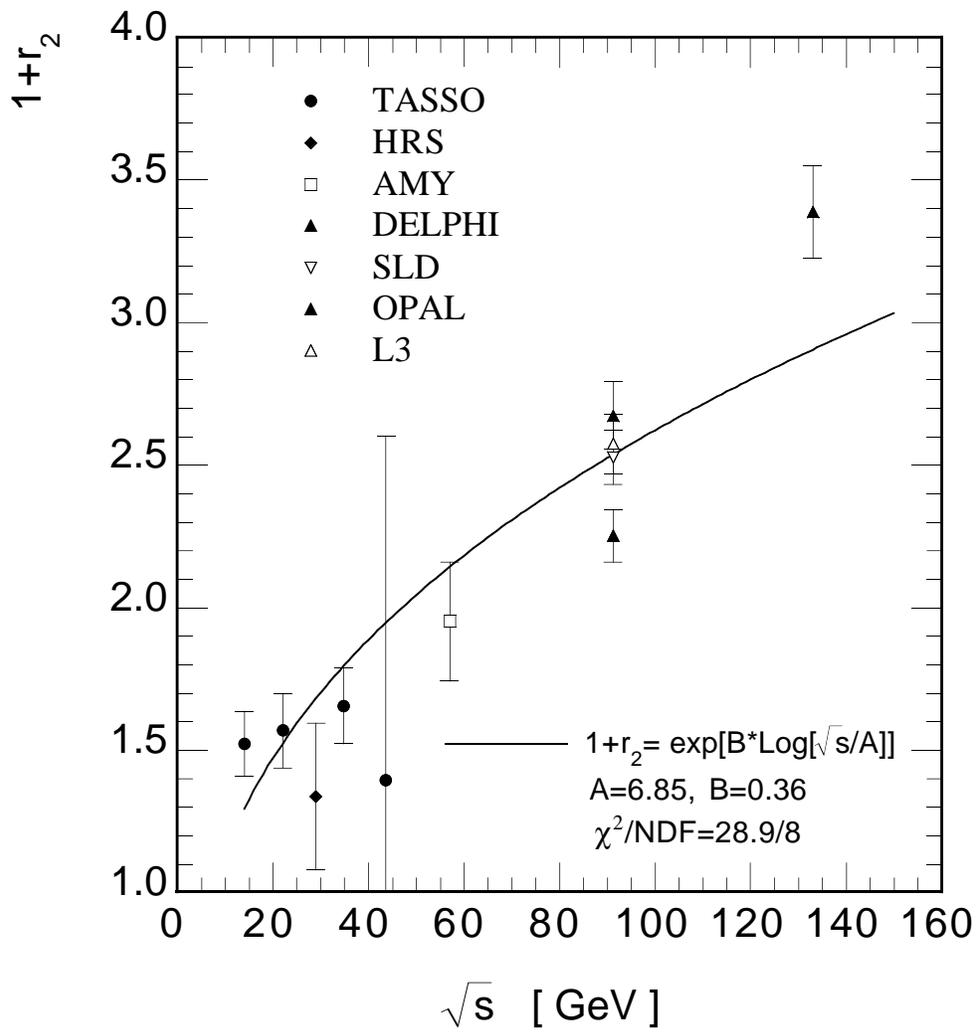
{\bf Fig.2:} \\
$ 1 + r_{2} = exp(\lambda t)$ vs  $\sqrt{s}$.  A function 
$ exp( \lambda ln(\sqrt{s}/s_0)) $ is chosen among 
various fittings by the MINUIT program: $\chi^2 = 28.9$.
We can estimate  $\lambda$ and $s_0$ in multiparticle dynamics.
 
 
\vspace*{2cm}
\newcommand{\lw}[1]{\smash{\lower2.0ex\hbox{#1}}}
\begin{center}
\begin{tabular}{|ll|r|lr||r|r|}
\hline
Exp. &[GeV]&$n_{max}$&\multicolumn{2}{c||}{[N]~~~MNBD}
&\multicolumn{1}{c|}{NBD}&\multicolumn{1}{c|}{Eq.(10)}\\
\hline \hline
TASSO &14. &26&[9~]&*9.34/10 &25.0/11&25.0/11  \\ \hline
TASSO &22. &28&[11]&*2.14/11 &10.1/12&10.3/11  \\ \hline
HRS   &29. &28&[20]&*6.24/11 &8.20/12&8.20/12  \\ \hline
TASSO &34.8&36&[13]&*11.8/15 &39.8/16&45.6/16  \\ \hline
TASSO &43.6&38&[38]& 7.42/16 &7.40/17& *6.82/17 \\ \hline
AMY   &57. &40&[12]&*9.84/17 &51.8/18&64.8/18  \\ \hline
DELPHI&91.2&52&[10]&*14.2/22 &185./23&278./23  \\ \hline
OPAL  &91.2&54&[9~]&*4.03/24 &47.5/25&76.2/25  \\ \hline
SLD   &91.2&50&[13]&*27.1/21 &154./22&301./22  \\ \hline
OPAL  &133.&54&[7~]&*4.57/21 &39.6/22&57.2/22  \\
\hline \hline
H1 & 80-115&18&[18]&  2.93/16& 2.92/17&*2.32/17  \\ \cline{2-7}
$1<\eta^{*}<5$
   &115-150&21&[21]& 8.52/19 & *7.01/20&7.80/20 \\ \cline{2-7}
   &150-185&22&[22]& 5.12/20 & 5.65/21&*4.75/21 \\ \cline{2-7}
   &185-220&23&[23]& 5.99/21 & 8.97/22&*5.53/22 \\ \cline{2-7}
\hline
\end{tabular}
\end{center}
\begin{center}
{\bf \large Table I}
\end{center}
 
\vspace*{1cm}
\begin{center}
\begin{tabular}{|ll|r|r|r|}
\hline
Exp. & [GeV] &  \multicolumn{1}{c|}{Eq.(7)}
&\multicolumn{1}{c|}{NBD}&\multicolumn{1}{c|}{Eq.(11)}\\
\hline\hline
TASSO & 14. & *10.9/11 & 19.0/12 & *11.0/12 \\  \hline
TASSO & 22. & *3.53/12 & 15.3/13 & *3.21/13 \\  \hline
HRS   & 29. &  15.0/12 & 37.4/13 & *11.9/13 \\  \hline
TASSO &34.8 & *8.85/16 & 42.7/17 & *9.23/17 \\  \hline
TASSO &43.6 &  14.1/17 & 49.9/18 & *12.3/18 \\  \hline
AMY   & 57  & *7.82/18 & 19.1/19 & 8.62/19 \\  \hline
DELPHI&91.2 & *17.1/23 & 21.0/24 & 141./24 \\  \hline
OPAL  &91.2 & *4.46/25 & *4.39/26& 29.9/26 \\  \hline
SLD   &91.2 & *35.3/22 & 200./23 & 85.0/23 \\  \hline
OPAL  &133  &  11.2/22 &*9.63/23 & 44.9/23 \\
\hline \hline
H1    &80-115& 20.5/17 & 37.7/18 &*18.9/18 \\ \cline{2-5}
$1<\eta^{*}<5$
      &115-150&19.5/20 &32.8/21 & *18.6/21 \\ \cline{2-5}
      &150-185& 9.55/21&18.4/22 & *9.04/22 \\ \cline{2-5}
      &185-220& 12.3/22&23.3/23 & *11.7/23 \\ \cline{2-5}
\hline
\end{tabular}
\end{center}
\begin{center}
{\bf \large Table II}
\end{center}
 
\vspace*{1cm}
 
\begin{table}[ht]
\centering
\begin{tabular}{|l|lccc|}\hline
Eq.(1) &P(n,t)& $\lambda$ & $\lambda_0$ & interval of $t$ \\
\hline
QCD &gluon dis.  & $\frac{N_c}{\varepsilon}$ &
$\frac{N_c^2-1}{2N_c\varepsilon}$ &
$\ln[\ln(Q^2/\Lambda^2)/\ln(Q_0^2/\Lambda^2)]$ \\
&&&&\\
multiparticle &\lw{hadron dis.}
&\lw{$\sim$0.36}& \lw {$0~(\mbox{or} k=\lambda_0/\lambda)$} &
\lw{$\ln(\sqrt{s}/6.85)$} \\
dynamics &&&&\\
\hline
\end{tabular}
\begin{center}
{\bf \large Table III}
\end{center}
\end{table}
\end{document}